\documentclass[reqno]{amsart}
\usepackage{graphicx,amsmath, amsaddr}
\usepackage{float}
\usepackage[utf8]{inputenc}
\usepackage{color}
\usepackage{cite}
\usepackage{fancyhdr}
\usepackage[margin=2cm]{geometry}
\usepackage{appendix}
\usepackage{multirow}
\usepackage[justification=centering]{caption}
\floatstyle{plaintop}
\restylefloat{table}
\usepackage{multirow}
\usepackage[table,xcdraw]{xcolor}
\usepackage{amssymb}
\usepackage{siunitx}
\usepackage{chemformula}
\usepackage{booktabs}
\usepackage{physics}
\title{Discovering correlations between metal foam thermal characteristics and non-Fourier behavior}

\author{A. Fehér$^{1}$, D. Cseh$^{1}$, R. Kovács$^{1,2}$}

\address{
	$^1$Department of Energy Engineering, Faculty of Mechanical Engineering, Budapest University of Technology and Economics, Műegyetem rkp. 3., H-1111 Budapest, Hungary \\
	$^2$Department of Theoretical Physics, Wigner Research Centre for Physics, Institute for Particle and Nuclear Physics, Budapest, Hungary
}
\date{\today}

\begin{document}

\maketitle

\begin{abstract}
Due to their low density and large specific surface area, metal foams are increasingly used as cellular materials that combine excellent structural and thermal properties. Their cellular structure makes them particularly suitable for use in heat exchangers, insulation, and fire protection layers. The heat transport that takes place within them is a complex phenomenon characterized by the simultaneous presence of heat conduction, heat transfer, and heat radiation, making their modeling a significant challenge. The aim of the research is to develop a one-dimensional, time-dependent, discrete numerical model capable of describing the effective thermal behavior of metal foams. The model takes into account heat conduction through the solid phase, conductive heat transfer in the closed cavities, thermal radiation between the pore walls, and by-passing heat conduction around the cavity. The results highlight that geometric features such as cavity size and arrangement have a significant impact on temperature distribution and confirm that classical Fourier-based models are not accurately applicable to porous materials, but the Guyer--Krumhansl equation is an adequate candidate.
%The developed model represents a good compromise between computational efficiency and physical realism and can serve as a useful tool for the design and optimization of metal foam-based thermal systems. 
Furthermore, we correlated the static and dynamic thermal diffusivity with the metal foam parameters, showing a significant sensitivity to the cavity thermal attributes.

\end{abstract}

\maketitle

\section{Introduction}
Porous, cellular metal structural materials (metal foams) have received increasing attention in engineering research and applications in recent decades. This is no coincidence, as these materials have significant mechanical strength despite their low density, while exhibiting special thermal, acoustic, and vibration-damping properties. Due to their internal complex geometry, metal foams are ideal for use in multifunctional structural solutions, such as in the automotive industry, aircraft manufacturing, or heat exchange systems \cite{ashby2001, femhab1, femhab2, femhab3, femhab4}. The thermal applications of metal foams are particularly promising, as they combine the good thermal conductivity of metals with the large specific surface area provided by the cell structure and the low thermal conductivity of the gas in the pores. As a result, metal foams can be used in thermal insulation, as a fire protection layer, and in compact heat exchangers where intensive heat transfer is required. Accurate modeling of the thermal behavior of metal foams is essential for the design and optimization of such systems. The Fourier heat conduction equation, which is applied to traditional, homogeneous materials, is unable to describe the heat transport processes occurring in heterogeneous, porous structures with sufficient accuracy. The presence of cells, their size, shape, and spatial arrangement can significantly modify the thermal conductivity characteristics, and the role of heat transfer and heat radiation cannot be ignored either. All this suggests that more complex, structure-sensitive models are needed for metal foams.

%\section{Fémhabok effektív tényezőjének meghatározása}

Accurate determination of the effective thermal conductivity of metal foams is crucial in their intended areas of application, particularly in thermal engineering systems. Several studies in the literature address this issue, using different experimental setups and measurement techniques.

Babcsán and co-workers \cite{babcsan2003} performed comparative thermal conductivity measurements on Alporas-type aluminum foams. The essence of the method is that the properties of a sample with unknown thermal conductivity were compared with those of known materials. 

%Három különböző relatív sűrűségű $(\rho_{rel} = \rho / \rho_0)$) mintát vizsgáltak, ahol $(\rho)$ a fémhab, $(\rho_0)$ pedig a tömör anyag sűrűsége. A méréseket 100$^\circ$ C, 200$^\circ$ C, 300$^\circ$ C és 400$^\circ$ C hőmérsékleteken végezték, egyes mintáknál vákuumban is. Az eredmények azt mutatták, hogy a vákuumban és légköri nyomáson mért értékek közel azonosak voltak (pl. (6.14 , W/mK) légköri nyomáson és (6.20, W/mK) vákuumban), így arra lehet következtetni, hogy az üregekben történő hőátadás és sugárzás 500$^\circ$ C alatt elhanyagolható.

Solórzano et al. \cite{solorzano2008} took a different approach, examining aluminum foams with different porosities using the Transient Plane Source (TPS) method. This technique allows for the spatial tracking of temperature changes over time, which is particularly useful for detecting local effects on the foam structure. By comparing the obtained measurement data with CT images, it was found that the internal inhomogeneity of the foam—for example, local variations in pore size and density—significantly affects the effective thermal conductivity. These results clearly demonstrate that the thermal properties of metal foams are highly dependent on their internal structure and that experimental studies provide an important basis for validating modeling approaches.

Numerous analytical models have been developed to determine the effective thermal conductivity of metal foams, which take into account the geometric and material characteristics of the cell structure to varying degrees. The aim of such models is to create an easily applicable relationship that allows for preliminary estimation of thermal behavior without experimental data. Among the simplest approaches are series and parallel thermal resistance models, which assume that heat conduction occurs through paths that are completely connected in series or in parallel between the solid phase and the porous (gas) phase. The series model underestimates the effective thermal conductivity, while the parallel model overestimates it. Although these approaches are easy to use, they are not suitable for describing the thermal conductivity behavior of real cell structures. Leach et al. \cite{leach1993} provide a detailed overview of analytical models, highlighting that most of them do not take into account either heat transfer in cavities or the role of thermal radiation. However, there are also models based on more realistic geometric idealizations. One such approach assumes that the structure of metal foam consists of regular square cells whose walls are made of matrix material. Another model assumes spherical pores, a geometry that better represents the internal structure of closed-cell foams. Boosma et al. \cite{boomsma2001} developed a one-dimensional analytical model based on a three-dimensional geometric structure that estimates the effective thermal conductivity of fluid-saturated metal foams. The model was based on a tetrakaidecahedron (14-sided polyhedron) cell structure, which closely approximates the cell structure formed during the manufacturing process.

%Az elmélet a szilárd és fluid fázisok térfogatarányát figyelembe véve képes előre jelezni a fémhab effektív hővezetését. A modell eredményei jól korrelálnak a mért adatokkal, és arra is rávilágítanak, hogy olyan esetekben, amikor a fém mátrix hővezető-képessége lényegesen magasabb, érdemes annak növelésével javítani a teljes szerkezet hőtechnikai teljesítményét.

%Ezen analitikus megközelítések hasznos első lépést jelenthetnek a fémhabok viselkedésének kvalitatív megértésében, azonban a komplex sejtszerkezetek és hőátadási mechanizmusok miatt gyakran csak korlátozott pontossággal alkalmazhatók konkrét tervezési feladatok során.

The complex, irregular cell structure of real metal foams differs significantly from the idealized geometric structures used in analytical models. As a result, the applicability of analytical approaches is limited, especially when high-precision results are required. Recognizing this problem, thermal conductivity studies based on numerical methods have become increasingly widespread in recent decades, especially in the case of metal foams. Computer simulations, such as the finite element method and the finite volume method, allow detailed thermal analyses to be performed even on foamed materials with complex inner structures. These methods are capable of taking into account the complexity of cell structures, geometric anisotropy, and spatial variations in material and thermal properties.

The first step in numerical modeling is always to accurately map the geometry of the foam. Some studies use 3D geometry based on image processing of real metal foam cross-sections, allowing them to create extremely realistic simulations. However, this is computationally intensive and not always suitable for quick engineering estimates or iterative design processes. For this reason, much research has focused on developing synthetic but statistically representative geometric models that strike the right balance between accuracy and computational cost.

Numerical simulations also allow us to examine the combined or separate effects of heat conduction, heat transfer, and heat radiation. In addition, parameter studies and sensitivity analyses can be performed to help map how porosity, cell size, material properties, or geometric arrangement affect effective thermal behavior. These methods are increasingly enabling the development of predictive models that not only serve research purposes but can also be applied in industrial-level design.

Artem and his colleagues \cite{artem2022} performed thermal testing of metal foams using the so-called digital twin method, which attempts to simulate the real process in as much detail as possible. During the experiments, several open-cell metal foams were analyzed by characterization, scanning, and measurements. The simulations and measurements yielded identical results, confirming the non-Fourier macroscopic thermal conductivity of metal foams. Earlier measurements conducted independently of Artem also showed the same properties.

However, determining the properties of a specific metal foam requires high-capacity calculations and simulations based on previous experience. Our goal, however, is to create a much more transparent, manageable, and faster model based on a simulation of the Fourier heat equation. We will show how the non-Fourier effect manifests itself in foam-like materials, and what material properties have a role in its formation, as well as how it may depend on these parameters, i.e., we will effectively characterize the transient behavior.

\section{Simulation preparation and characteristics}

\begin{figure}[H]
    \centering
    \includegraphics[width=0.5\linewidth]{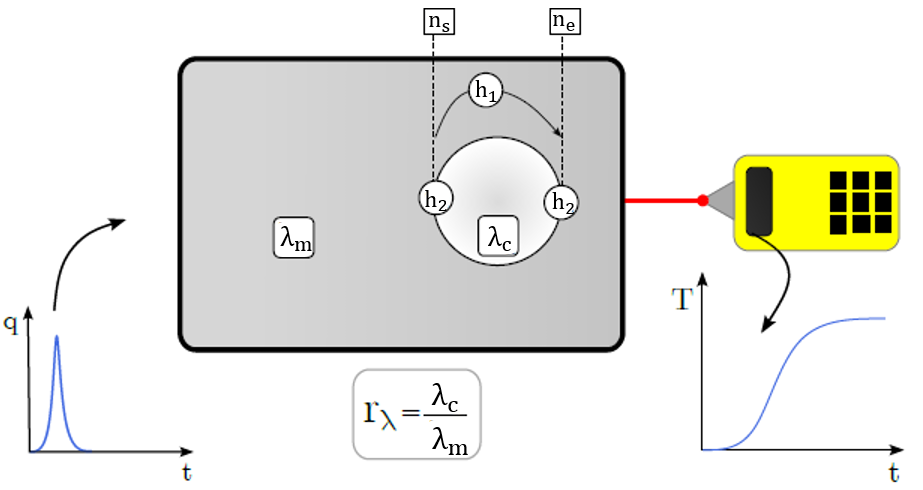}
    \caption{The model's structure, where $\lambda_m$ is the thermal conductivity of the metal, $\lambda_c$ is the thermal conductivity of the cavity; $n_s$ and $n_e$ are the beginning and the end of the cavity.}
    \label{fig:model}
\end{figure}

Our aim is to prepare a foam-like model, which we can call a single cavity model. This keeps the approach as straightforward as possible, but also keeps the computing needs manageable. We chose the heat pulse experiment for our virtual experiment due to the available experience in the literature \cite{feh24femhab} showing thermal behavior beyond Fourier. We model the heat pulse experiment by applying a transient heating on the left boundary to model the time variation of the temperature on the right boundary, as shown in Figure~\ref{fig:model}. As an initial condition for the simulation, it is assumed that the temperature of the body is equal to the ambient temperature at the start, and therefore, the heat flux density is homogeneously zero throughout the body.

\begin{align}
    T(t=0,x) = T_0, \nonumber \\
    q(t=0,x) = 0. \nonumber
\end{align}

On the left side, as shown in Figure \ref{fig:model}, a time-dependent heat flux excitation is prescribed.

\begin{align}
    q(t, x=0) =
\begin{cases}
q_{\text{max}} \left( 1 - \cos\left( 2\pi \frac{t}{t_p} \right) \right), & \text{if } t < t_p, \\
0, & \text{if } t > t_p,
\end{cases}
\end{align}

It is sufficient to define the internal energy balance equation to describe heat conduction in the bulk, but we must take into account not only heat conduction but also heat transfer and (linearized) thermal radiation in the cavity when setting up the model.

The parameters introduced, which are illustrated in Figure \ref{fig:model}, are as follows:

\begin{itemize}
    \item $h_1$: equivalent heat transfer coefficient, which also models the heat radiation between the two walls of the cavity and the heat conduction in the metal surrounding the cavity.
    \item $h_2$: heat transfer coefficient, which models the heat transfer coefficient between the cavity wall and the medium inside the cavity.
    \item $r_\lambda$: dimensionless parameter, the medium in the cavity, and the thermal conductivity of the metal.
    \item $n_s$: taking into account the position of the cavity. We keep the cavity in the same size over the entire study.
\end{itemize}
Later, when we create data sets as virtual experiments, $h_2$ is kept unchanged since the cavity is closed, no substantial fluid flow effects are assumed due to the slight temperature increase in real experiments (about $3-5$ K). Furthermore, we assumed that linearizing the thermal radiation heat flux is sufficient and built it into $h_1$, in the same way as we assume bypass heat conduction around the cavity. 

\subsection{Dimensionless parameters and equations}
We introduce dimensionless quantities, thus obtaining relationships applicable to general cases. The dimensionless parameters are:
\begin{align}
&\textrm{time and spatial coordinates}: & Fo =\frac{\alpha_{stat}}{L^2}t, \quad &  \textrm{and}  \quad \xi=\frac{x}{L}; \nonumber \\
&\textrm{} \\
&\textrm{temperature:} & \vartheta=\frac{T-T_{0}}{T_{\textrm{end}}-T_{0}} \quad &\textrm{with} \quad
T_{\textrm{end}}=T_{0}+\frac{\bar{q}_0 t_p}{\rho c L}; \nonumber \\
\label{ndvar}
\end{align}
where $\alpha_{stat}=\lambda/(\rho c)$ is the thermal diffusivity, $Fo$ is the Fourier number and $\bar{q}_0=\frac{1}{t_p}  \int_{0}^{t_p} q_{0}(t)\textrm{d}t$.

By rearranging the dimensionless quantities and then substituting them into the equations used, the dimensionless form is obtained:
\begin{align}
    &\textrm{Fourier's law:} &q = - \frac{\partial\vartheta}{\partial \xi}, \nonumber\\
    &\textrm{heat pulse boundary:} &q= \bigg(1- \cos \bigg(2 \pi\frac{Fo}{Fo_p}\bigg)\bigg), \nonumber\\
    &\textrm{energy balance equation:} &\frac{\partial\vartheta}{\partial Fo} + \frac{\partial q}{\partial \xi} = 0. \nonumber
\end{align}
\subsection{Discretization}
\begin{figure}[H]
    \centering
    \includegraphics[width=0.75\linewidth]{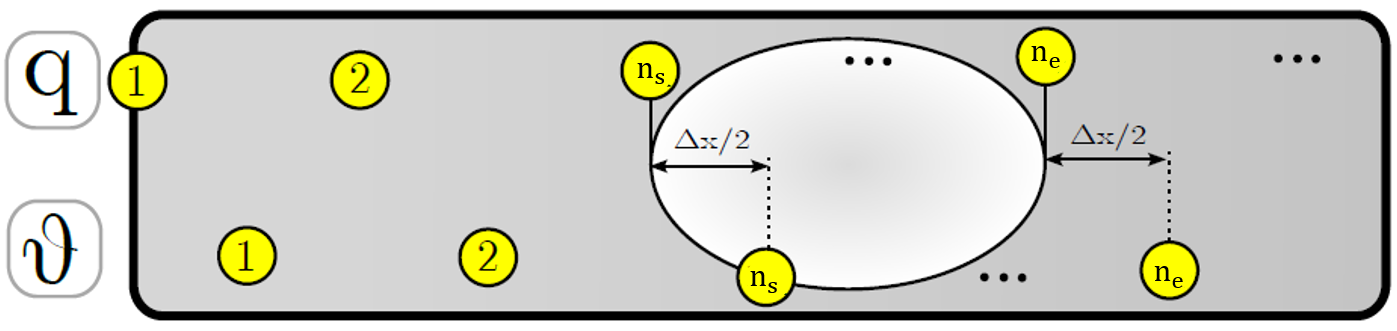}
    \caption{Illustration of shifted field discretization.}
    \label{fig:discretization}
\end{figure}

To solve the equations, we use an explicit staggered field discretization, see Figure \ref{fig:discretization} for details. This means that the heat fluxes are calculated at the boundaries of each cell, and the temperature is calculated only inside a cell. The discretized version of the equations, where $j$ denotes the time step and $n$ denotes the spatial step, reads
\begin{align}
    &\textrm{Fourier law}: &q^{j+1}_n = - \frac{\vartheta^{j+1}_{n+1/2} - \vartheta^{j+1}_{n-1/2}}{\Delta \xi}, \label{eq:disc1} \\
    &\textrm{energy balance equation:} & \vartheta^{j+1}_{n+1/2} =  \vartheta^{j}_{n-1/2} - \frac{\Delta Fo}{\Delta \xi} (q^{j}_{n+1} - q^{j}_{n}) \label{eq:disc2}.
\end{align}
When there are cavities in the material, the equations also change. The form we use depends on which point in space we are calculating the heat flow for.

The heat flux point at the beginning of the cavity is denoted by $n_s$, and the heat flux point at the end is denoted by $n_e$. The equations change in different domains:
\begin{itemize}
    \item if $n < n_s$ and $n > n_e$, the equations are formed according to \eqref{eq:disc1} and \eqref{eq:disc2};
    \item if $n = n_s$ or $n = n_e$, then the heat flow at the point is calculated only taking into account heat transfer and heat radiation at the boundary using the $q^{j+1}_{n_s}=h_1 (T^{j+1}_{n_s}-T^{j+1}_{n_e})$ formula;
    \item if $n_s < n < n_e$, then the heat flow is calculated taking into account the heat conduction in the cavity with a reduced thermal conductivity, introducing $r_\lambda$ for the heat flux: $q \rightarrow r_\lambda q$.
\end{itemize}
During the simulation, we also performed a grid independence test, which showed that above 150 cells, the difference between the rear side temperatures disappears, so we will perform subsequent simulations with that grid. Furthermore, we use a 10-cell-wide cavity, which always fixes the value if $n_e$.

\subsection{Sensitivity test}

We examine the dependence of the non-Fourier effect on various thermal parameters using a sensitivity analysis method, which shows how the output of a given model changes when the input parameters are varied. The analysis was performed based on the article \cite{feh24femhab}. Essentially, the simulation runs with a small change in a given parameter, and then the difference between the two data sets is divided by the difference in parameters to obtain the sensitivity function:
\begin{align}
    S_{p_i, t} = \frac{\partial y}{\partial p_i}, \quad y = y(p_i, t), \quad i = 1,...,n, 
\end{align}
where $y(p_i,t)$ is a function for which sensitivity is examined, in our case, the heat transfer parameters such as $h_1$, $r_\lambda$, and the position and size of the cavity.

In order to ensure that the sensitivity functions obtained are comparable with each other, it is necessary to prepare the so-called reduced sensitivity functions:

\begin{align}
     \hat{S}_{p_i, t} = p_i \frac{\partial y}{\partial p_i},  \quad y = y(p_i, t), \quad i = 1,...,n.
\end{align}

In the case of sensitivity functions $h_1$, it can be seen that the model is most sensitive in the 0-1 dimensionless time range, and as time progresses, the parameter no longer influences the results, i.e., it has no effect on the equilibrium temperature value. The reduced sensitivity functions clearly show that the size of the cavity has the greatest effect on the model output, while the position of the cavity has a smaller effect on the result.

\begin{figure}[H]
 \centering
    \includegraphics[width=0.45\linewidth]{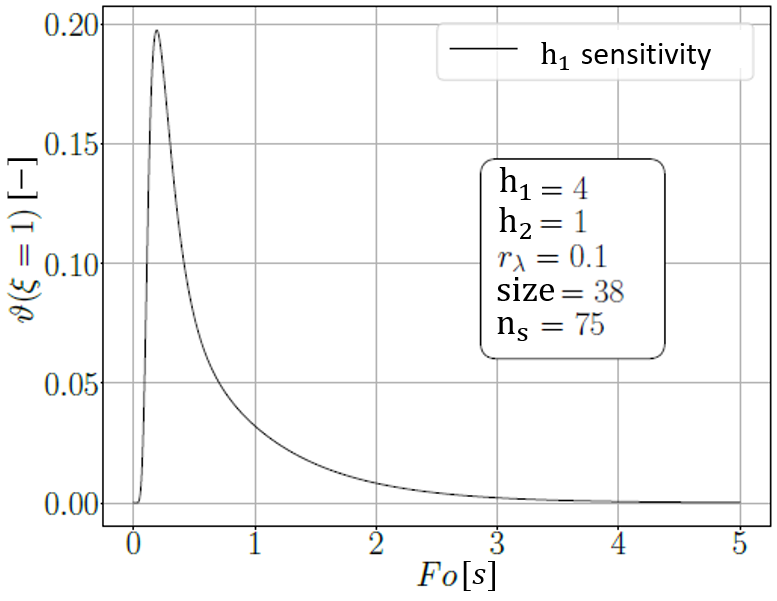}
    \includegraphics[width=0.45\linewidth]{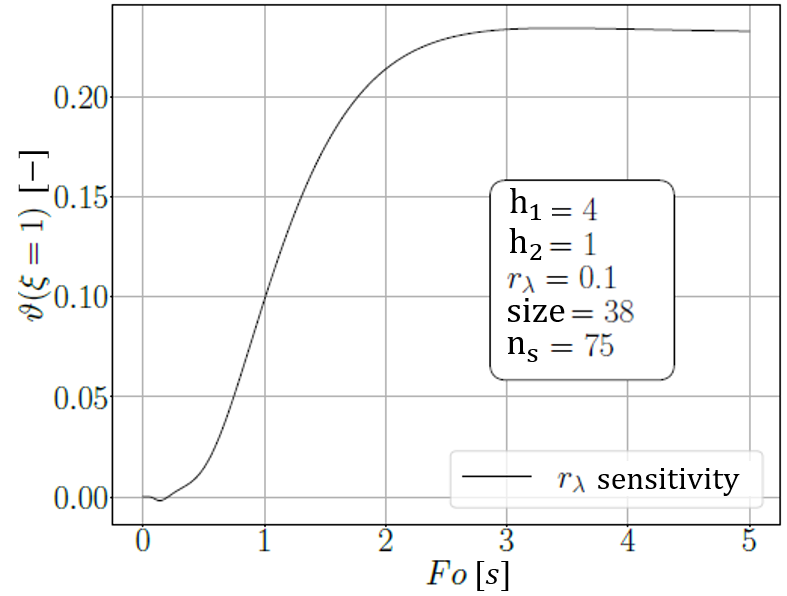}
    \includegraphics[width=0.45\linewidth]{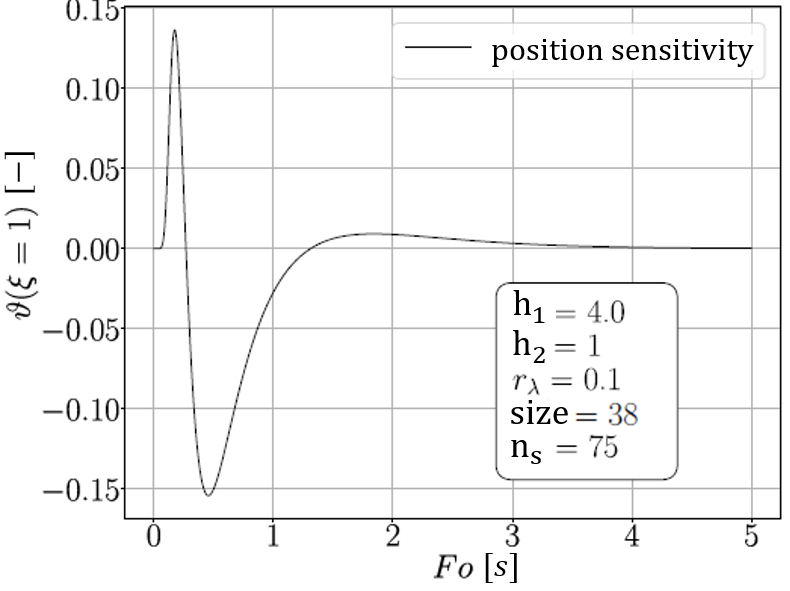}
    \includegraphics[width=0.45\linewidth]{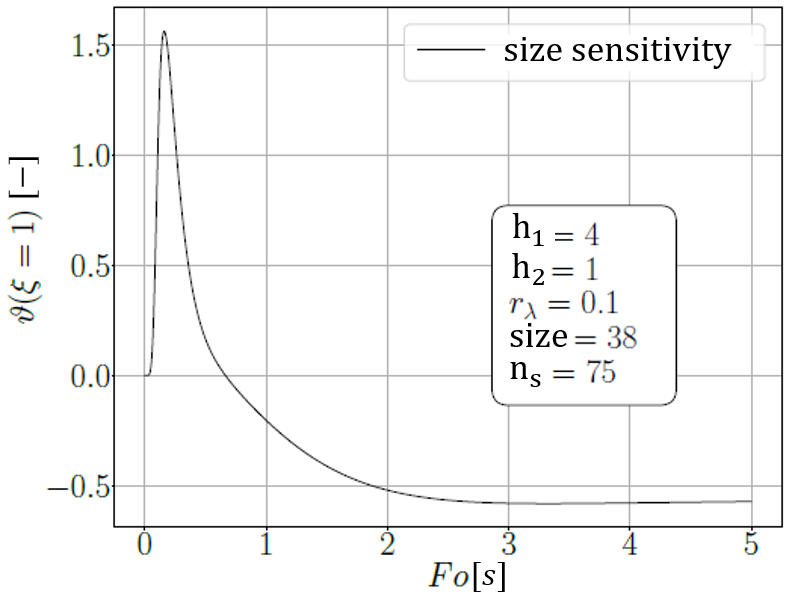}
    \caption{Reduced sensitivity functions of parameters as a function of time.}
\label{fig:erzekenyseg}
\end{figure}

\subsection{Data post-processing}

As a result of the simulations, the data sets obtained provide a virtual data set for a specific heat pulse experiment, with the difference that the measurements are not affected by noise. This is particularly important for our purposes, as it gives the simulation a clean output with no cooling, so that the equilibrium temperature is present at a dimensionless temperature of 1, while the initial temperature is 0, which can then be fitted using various models.

When creating virtual data sets, it is possible to specify the properties of heterogeneous materials and then examine how certain properties influence the static and dynamic thermal diffusivity values. Using the model, it is possible to examine the extent to which the parameter values are influenced by different $r_\lambda$ values, cavities of different sizes, different cavity positions, and changes in the $h_1$ bypass thermal conductivity value.

\section{Introduction to the effective model}

\subsection{Effective models}
After creating the data sets, one and two time scale models are used to determine the material parameters by fitting the simulation like a virtual experiment using the method described in \cite{fehkov24}. The $T$-representation of the Fourier equation:
\begin{align}
    \partial_tT = \alpha_{stat}  \partial_{xx}T,
\end{align}
where $\alpha_{stat}$ is the static thermal diffusivity, which characterizes processes with a single time scale. However, in the case of transient processes related to materials with complex material structure, a single time scale and a single thermal diffusivity might not be sufficient to characterize them. To evaluate the data, the Guyer--Krumhansl equation is used as an alternative two time scale equation, which, in addition to the static thermal diffusivity  $\alpha_{stat}$, also includes a dynamic coefficient, which can be expressed by the ratio $\frac{\kappa^2}{\tau}$ in the equation,
\begin{align}
    \tau \partial_{tt}T + \partial_tT = \alpha_{stat} \partial_{xx}T + \kappa^2 \partial_{xxt}T.
\end{align}

Previous experiments have already proven the existence of static and dynamic thermal diffusivity and their variation for different metal foam characteristics. This article examines clean, noise-free data sets obtained through simulation, but in the case of simulations, it is known which parameter changes were used to obtain the data set under examination. It is our aim to examine how changes in certain material parameters affect thermal diffusivity. The material parameters we observed changes in were: $r_\lambda$, cavity position, and the $h_1$. The evaluation was performed using an iterative method based on sensitivity analysis, the details of which can be found in \cite{fehkov24}.

\subsection{The effect of changing the parameter $r_\lambda$}

\begin{figure}[]
    \centering
    \includegraphics[width=0.45\linewidth]{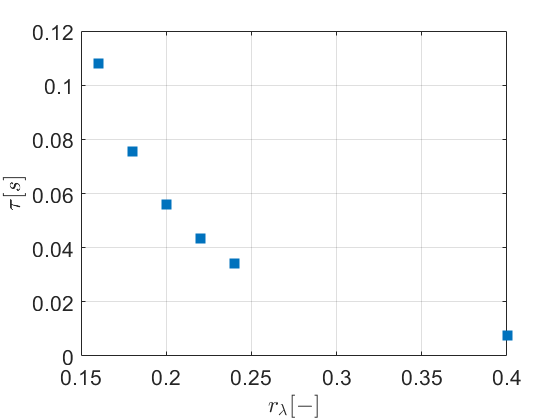}
    \includegraphics[width=0.45\linewidth]{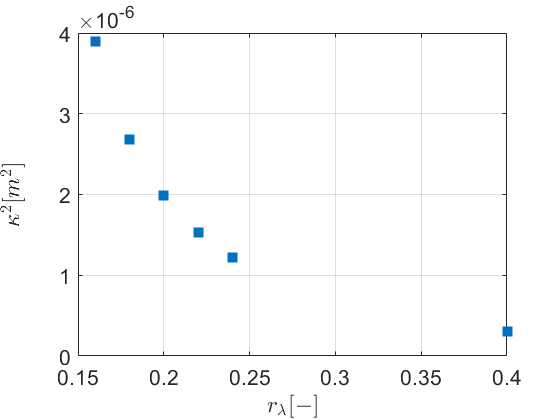}    \includegraphics[width=0.45\linewidth]{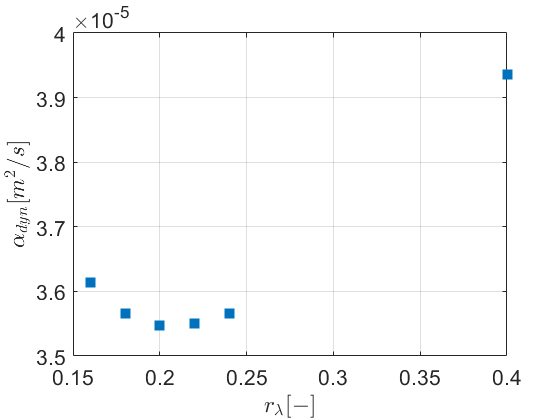}
    \includegraphics[width=0.45\linewidth]{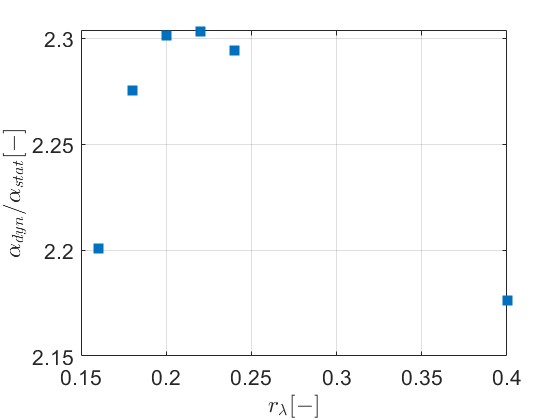}
    \caption{Changes in various parameters ($\tau, \kappa^2, \alpha_{dyn}, \alpha_{dyn}/\alpha_{stat}$) as a result of increasing $r_\lambda$.}
    \label{fig:rlvalt}
\end{figure}

\begin{figure}[]
    \centering
    \includegraphics[width=0.45\linewidth]{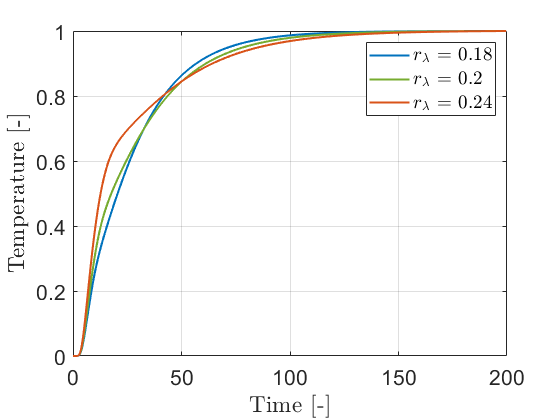}
    \includegraphics[width=0.45\linewidth]{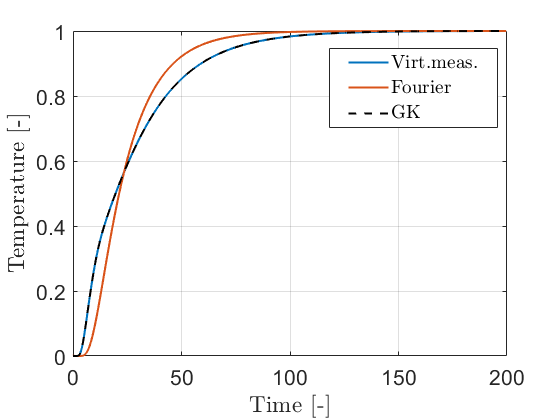}
    \caption{Left: Change in temperature profile as a result of changing the parameter $r_\lambda$. Right: Fitting of the virtual measurement temperature profile using the Fourier and GK equations.}
    \label{fig:rlambda_plot}
\end{figure}

By varying the value of $r_\lambda$ between 0.18 and 0.4, it can be observed that smaller values significantly influence the non-Fourier parameters, which form the dynamic thermal diffusivity. It is also insightful to see how the ratio of static and dynamic thermal diffusivity behaves. If their ratio equals 1, it would lead to the Fourier equation since both time scales coincide. In our virtual experiments, we observed a non-monotonous behavior in $r_\lambda$, demonstrated in Figure \ref{fig:rlvalt}. When the matrix and cavity fluid thermal conductivity greatly differ, the most influential factor is the difference between the two time scales in the heat transfer process. This can be characterized effectively with the Guyer--Krumhansl heat equation, see Figure~\ref{fig:rlambda_plot}. Increasing $r_\lambda$ results in a closer gap between the conduction time scales; therefore, the process approaches the single time scale behavior. 

\subsection{The effect of changing the cavity position}
\begin{figure}[]
    \centering
    \includegraphics[width=0.45\linewidth]{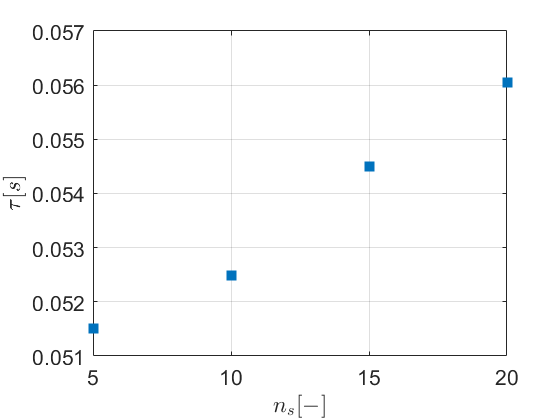}
    \includegraphics[width=0.45\linewidth]{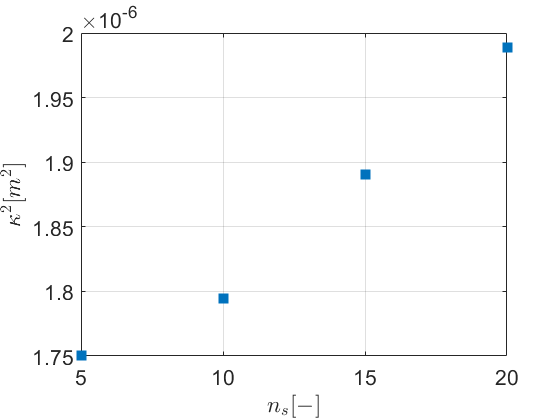}    \includegraphics[width=0.45\linewidth]{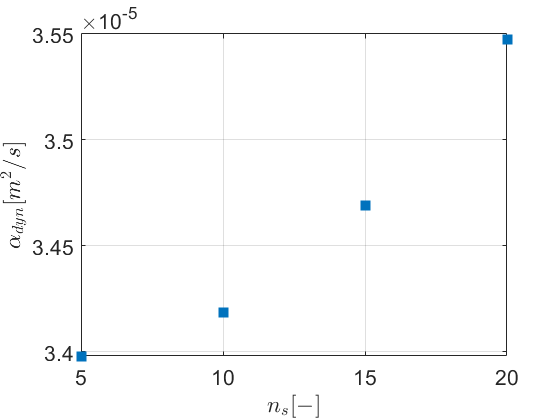}
    \includegraphics[width=0.45\linewidth]{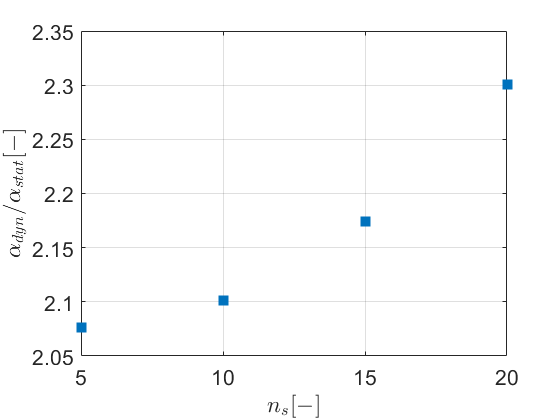}
    \caption{Changes in various parameters ($\tau, \kappa^2, \alpha_{dyn}, \alpha_{dyn}/\alpha_{stat}$) as a result of changing the cavity position.}
    \label{fig:mvalt}
\end{figure}

\begin{figure}[]
    \centering
    \includegraphics[width=0.45\linewidth]{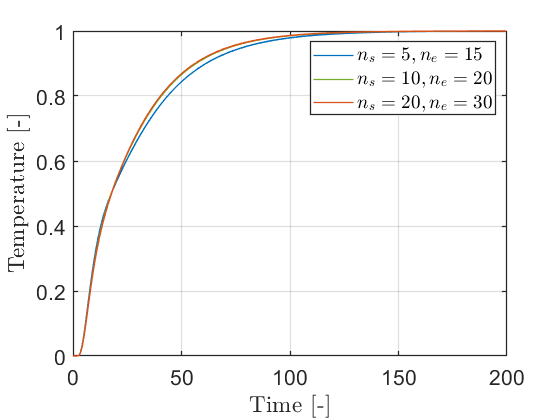}
    \includegraphics[width=0.45\linewidth]{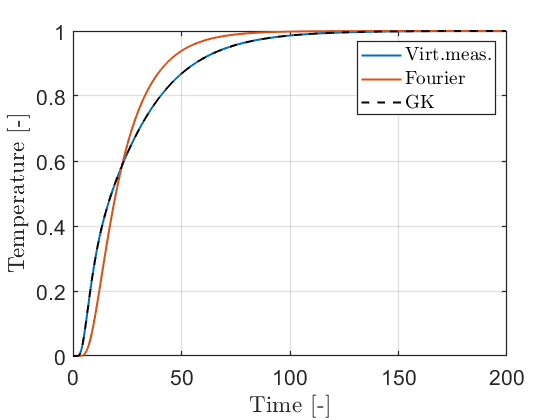}
    \caption{Left: Change in temperature profile as a result of changing the cavity position. Right: Fitting of the virtual measurement temperature profile using the Fourier and GK equations.}
    \label{fig:m_plot}
\end{figure}

By changing the position of the cavity within the material, it was examined how it changes the thermal response when located close to the heat pulse, in the middle of the material, and at the end of the simulated material. Based on the observations, it can be stated that the position of the cavity affects the thermal characteristics of the material in a monotonous way. When the cavity is placed closer to the end of the body, there is less time for the temperature to equilibrate and thus eliminate the differences originating by the presence of two conduction time scales. This is observable in the ratio of the static and dynamic thermal diffusivity (Figure~\ref{fig:mvalt}). Figure \ref{fig:m_plot} demonstrates the change in the rear side temperature history.

\subsection{The effect of changing the $h_1$}
\begin{figure}[]
    \centering
    \includegraphics[width=0.45\linewidth]{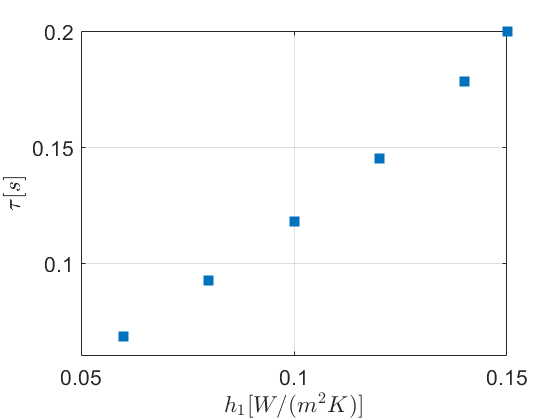}
    \includegraphics[width=0.45\linewidth]{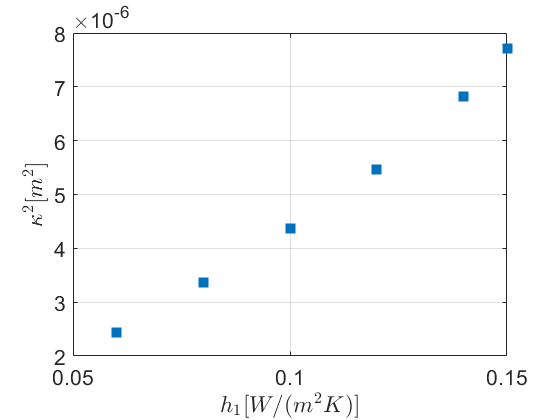}    \includegraphics[width=0.45\linewidth]{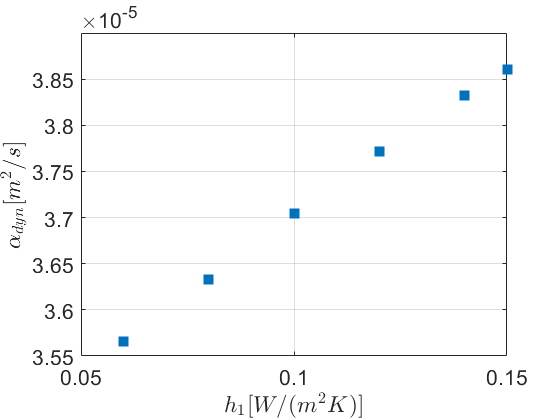}
    \includegraphics[width=0.45\linewidth]{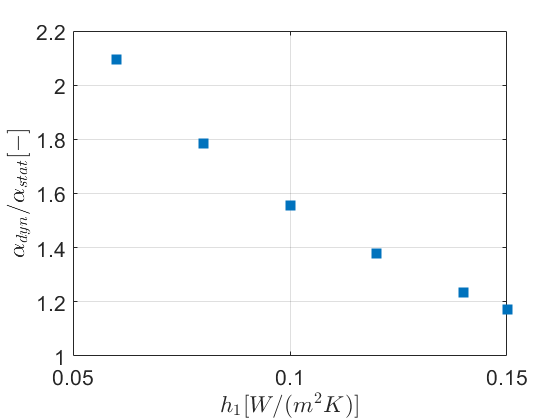}
    \caption{Changes in various parameters ($\tau, \kappa^2, \alpha_{dyn}, \alpha_{dyn}/\alpha_{stat}$) as a result of increasing $h_1$.}
    \label{fig:hvalt}
\end{figure}

\begin{figure}[]
    \centering
    \includegraphics[width=0.45\linewidth]{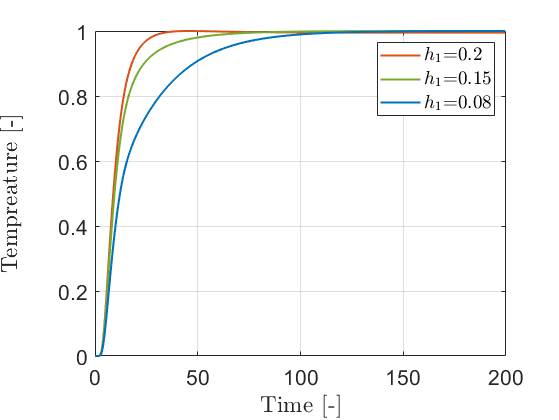}
    \includegraphics[width=0.45\linewidth]{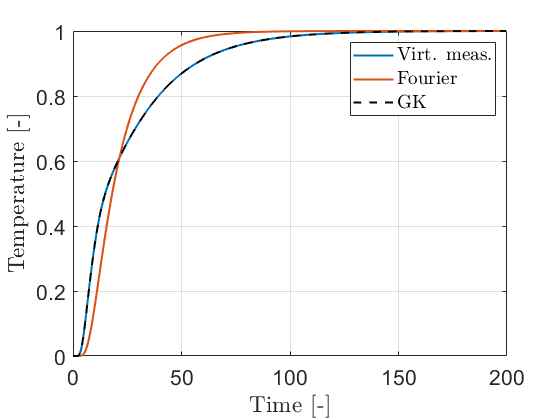}
    \caption{Left: Change in temperature profile as a result of changing the $h_1$. Right: Fitting of the virtual measurement temperature profile using the Fourier and GK equations.}
    \label{fig:h_plot}
\end{figure}

By varying the cavity size in the range of 0.08-0.15, we found that it significantly changes the thermal response, so the non-Fourier parameters also change, increasing with the increase in cavity size. Although there is a monotonous change in the dynamic thermal parameters, the static thermal diffusivity increases even faster, leading to a significant decrease in the ratio of static and dynamic thermal diffusivity, and hence their ratio quickly approaches 1, the single time scale case, demonstrated in Figure \ref{fig:hvalt}. Figure \ref{fig:h_plot} shows the change in the observable rear side temperature history.

\section{Summary and discussion}

In the present paper, we introduced a simple but efficient model for the characterization of a foam-like material including a single cavity. We have taken into account the heat conduction and heat convection effects inside the cavity. Additionally, we also modeled the effects of (linearized) radiation and bypass heat conduction using a separate heat transfer coefficient. This approach enabled us to investigate the cavity-related thermal parameters and their effects on the effective thermal diffusivity. Therefore, we introduced three parameters for which we performed multiple simulations, and treated them like a virtual experiment. 

These virtual experiments strengthen the earlier real experimental evidence: the presence of multiple heat transfer channels can lead to a thermal history that cannot be modeled with the Fourier heat equation, as it is limited to a single time scale process. The Guyer--Krumhansl equation seemed an excellent alternative to the Fourier equation due to the earlier experiments, and because of the presence of two time scales, characterized by the static and dynamic thermal conductivity. As it is expected, the more homogeneous the material, the closer we can find the rear side temperature history to a single time scale process. However, when a cavity-related thermal parameters significantly differ from the matrix material, the presence of two distinct heat transfer channels starts to dominate the initial transients. These cases demand a two time scale model to describe, and this is not restricted to the Guyer--Krumhansl heat equation. 

The Jeffreys heat equation might also be a good candidate due to its same $T$-representation as the Guyer--Krumhansl heat equation, and thermodynamic compatibility. The linear version of the dual-phase-lag model can be considered as well, but we wish to highlight its mathematical and physical issues, indicating its strong limitations towards real engineering applications.

\section*{Acknowledgements}
The research reported in this paper and carried out at BME has been supported by the grants of the Ministry for Innovation and Technology project EKÖP-24-3-BME-67. The research was funded by the Sustainable Development and Technologies National Programme of the Hungarian Academy of Sciences (FFT NP FTA).

\section*{Author contributions}
A.F.: model preparation, simulations, parameter fitting, manuscript preparation. D.Cs.: simulations, coding. R.K.: conceptualization, funding, manuscript preparation, supervision.

\bibliographystyle{unsrt}
%\bibliography{references}

\begin{thebibliography}{20}


\bibitem{ashby2001}
M.F. Ashby, A. Evans, N.A. Fleck, L.J. Gibson, J.W. Hutchinson, H.N.G. Wadley, and F. Delale. Metal Foams: A Design Guide. Applied Mechanics Reviews, 54(6):B105–B106, 11 2001.

\bibitem{femhab1}
K. Boomsma, D. Poulikakos, and F. Zwick. Metal foams as compact high performance
heat exchangers. Mechanics of Materials, 35(12):1161–1176, 2003.

\bibitem{femhab2}
Hamid Salimi Jazi, Javad Mostaghimi, L. Pershin, and Tom Coyle. Spray-formed,
metal-foam heat exchangers for high temperature applications. Journal of Thermal
Science and Engineering Applications, 1, 09 2009.

\bibitem{femhab3}
John Banhart. Manufacture, characterisation and application of cellular metals and metal foams. Progress in Materials Science, 46(6):559–632, 2001.

\bibitem{femhab4}
Francisco García-Moreno. Commercial applications of metal foams: Their properties and production. Materials, 9(2), 2016.

\bibitem{babcsan2003}
N. Babcsán, I. Mészáros, and N. Hegman. Thermal and electrical conductivity measurements on aluminum foams. Materialwissenschaft und Werkstofftechnik,
34(4):391–394, 2003.

\bibitem{solorzano2008}
E. Solórzano, J.A. Reglero, M.A. Rodríguez-Pérez, D. Lehmhus, M. Wichmann, and J.A. de Saja. An experimental study on the thermal conductivity of aluminium foams by using the transient plane source method. International Journal of Heat and Mass
Transfer, 51(25-26):6259–6267, 2008.

\bibitem{leach1993}
A. G. Leach. The thermal conductivity of foams. i. models for heat conduction. Journal of Physics D: Applied Physics, 26(5):733–739, 1993.

\bibitem{boomsma2001}
K. Boomsma and D. Poulikakos. On the effective thermal conductivity of a threedimensionally structured fluid-saturated metal foam. International Journal of Heat and Mass Transfer, 44(4):827–836, 2001.

\bibitem{artem2022}
A. Lunev, A. Lauerer, V. Zborovskii, and F. Léonard, Digital twin of a laser flash experiment helps to assess the thermal
performance of metal foams, Int. J. Therm. Sci., vol. 181, p. 107743, 2022.

\bibitem{feh24femhab}
Anna Fehér, J. E. Maróti, D. M. Takács, I. N. Orbulov, and R. Kovács. Thermal and mechanical properties of AlSi7Mg matrix syntactic foams reinforced by Al2O3 or SiC particles in matrix. International Journal of Heat and Mass Transfer, 226:125446, 2024.

\bibitem{fehkov24}
A. Fehér and R. Kovács. On the dynamic thermal conductivity and diffusivity observed in heat pulse experiments. Journal of Non-Equilibrium Thermodynamics, 49:2 pp. 161-170., 10 p., 2024.


\end{thebibliography}

\end{document}